\documentstyle[twoside,fleqn,npb,epsfig]{article}
\newcommand{\lum}{${\cal L}$}
\newcommand{\piz}{${\pi^{\circ}}$}
\newcommand{\ptpi}{${p_{T,\pi}}$}
\newcommand{\etapi}{${\eta_{\pi}}$}
\newcommand{\thpi}{${\theta_{\pi}}$}
\newcommand{\pt}{${p_{T}}$}
\newcommand{\xpi}{${x_{\pi}}$}
\newcommand{\epi}{${E_{\pi}}$}
\newcommand{\eprot}{${E_{\rm proton}}$}
\newcommand{\xbj}{${x}$}
\newcommand{\ybj}{${y}$}
\newcommand{\qsq}{${Q^2}$}
\newcommand{\gevsq}{${\mathrm{GeV}^2}$}

\newcommand{\AmS}{{\protect\the\textfont2
  A\kern-.1667em\lower.5ex\hbox{M}\kern-.125emS}}

\title{Forward {\piz}-meson production at HERA }

\author{T. Wengler 
  \address{Physikalisches Institut, Universit\"at
    Heidelberg, Philosophenweg 12, 69120 Heidelberg, Germany}
  -- talk given on 
behalf of the H1 collaboration at DIS99, Zeuthen}

\begin{document}

\begin{abstract}
 The production of high transverse momentum {\piz}-mesons has been measured
 in deep-inelastic e-p scattering events at
 low Bjorken-$x$ taken with the H1 detector at HERA. The production
 of high {\pt} particles is strongly correlated to the emission of
 hard partons in QCD and is therefore sensitive to the dynamics of
 the strong interaction. For the first time the measurement of single
 particles has been extended to the region of small angles w.r.t. the
 proton remnant (forward region) and down to very low values of
 $x \approx 5{\cdot}10^{-5}$. This region is expected to be
 particularly sensitive to QCD evolution effects in final
 states. Differential cross sections
 of inclusive {\piz}-meson production have been measured as a function
 of Bjorken-$x$ and 
 the four-momentum transfer {\qsq}, and also as a function of
 the transverse momentum and the polar angle of the {\piz}-mesons.
 A recent BFKL calculation and QCD models based on the DGLAP splitting
 functions are compared to the data. The best description of the data
 is achieved by the BFKL calculation.
\end{abstract}

\maketitle

\section{INTRODUCTION}
It is the unique kinematical reach of the ep collider HERA which has
enabled us to study deep-inelastic scattering (DIS) at values of Bjorken-$x$
down to $x \sim 10^{-6}$ as well as at momentum transfers up to
{\qsq} $\sim$ 30000~{GeV$^2$}. 

In the classical DIS picture
a parton in the proton can undergo a QCD cascade resulting in several 
parton emissions before the final parton interacts with the virtual
photon. 
Differences between different dynamical assumptions on the parton cascade
are expected to be emphasized in the region towards the proton remnant
direction, i.e. away from the scattered quark in the HERA kinematical
range. In the HERA laboratory frame 
this has been termed the forward region.

In this paper we study forward single {\piz} production for a
considerably larger data sample and 
in an enlarged kinematical range as compared to a previous publication
by the  H1 collaboration~\cite{h1fwdjet}.
The production of high {\pt} particles is strongly correlated to the
emission of hard partons in QCD~\cite{twthesis} and is therefore
sensitive to the dynamics of the strong interaction.

\section{MEASUREMENT}

\begin{figure}[t]
\begin{center}
\includegraphics[width=0.4\textwidth,height=10cm,trim=0 50 0 15]
{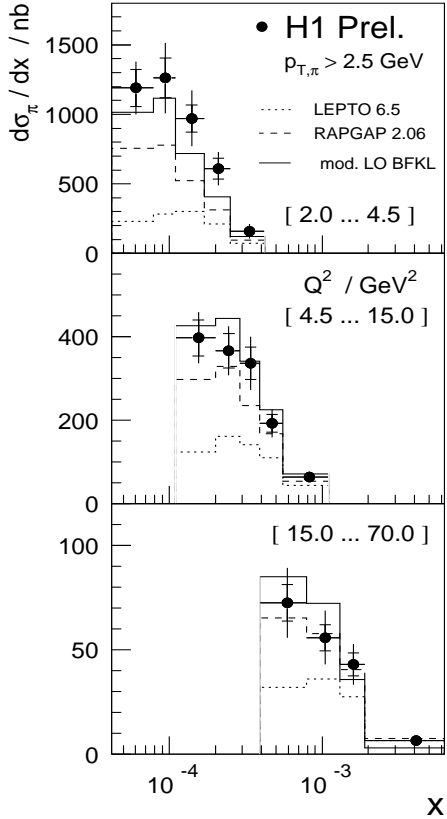}
\end{center}
\caption{\it Inclusive {\piz}-meson cross sections as a
  function of {\xbj} for {\ptpi}(hcms) $> 2.5$~GeV, 0.1 $<$ {\ybj} $<$
  0.6, $5^\circ <$ {\thpi} $< 25^\circ$ and {\xpi} $=$
  {\epi}$/${\eprot} $>$ 0.01.
  RAPGAP (dir+res) and LEPTO
  are compared to the data as well as the prediction of the modified LO
  BFKL calculation~\cite{KwMaOu}}
\label{fig:xbj}
\vspace*{-0.5cm}
\end{figure}

The analysis is based on data representing an integrated luminosity of
{\lum}$ = 5.8~{\rm pb}^{-1}$ taken by H1 during the 1996 running period.
Deep-inelastic scattering events are selected in the range $0.1 < y <
0.6$ and $2 < Q^2 < 70$~GeV$^2$. About 600k events remain
after the selection. 

The {\piz}-mesons are measured using the dominant decay channel 
{\piz} $\rightarrow 2\gamma$. The {\piz} candidates are selected in
the region  $5^{\circ} < ${\thpi}$ < 25^{\circ}$, where {\thpi} is the polar
angle of the produced {\piz}. Candidates are required to have an energy of 
{\xpi }$=${\epi}/{\eprot} $>$ 0.01, with {\eprot} the proton beam energy, and 
a transverse momentum in the hadronic cms, {\ptpi}, greater than  2.5~GeV.
At the high {\piz} energies considered here, the two photons from the
decay  cannot be separated, but appear as one object (cluster) in the
calorimetric response. The standard method of reconstructing the
invariant mass from the separate measurement of the two decay photons
to identify the {\piz}-meson is hence not applicable.
Instead, a detailed analysis of the longitudinal and transverse shape
of the energy depositions is performed~\cite{twthesis}. 
This approach is based on the
compact nature of electromagnetic showers as opposed to showers of
hadronic origin, which are broader. 
The main experimental challenge in this analysis is the
high particle and energy density in this region of phase space, with
hadronic showers 
`obscuring' the clear electromagnetic signature provided by the two
photons of a {\piz} decay. This overlap is mainly responsible for losses
of {\piz} detection efficiency, since the distortion of the shape
estimators it causes will in many cases lead to the rejection of the
cluster candidate.
With this selection about 1700 {\piz} candidates are found  with a
detection  efficiency above 45$\%$. 

Monte Carlo studies using a detailed simulation of the H1 detector
yield a purity of about 70\% for the selected {\piz}-meson sample,
with impurities from misidentified hadrons from the main vertex, partly
(10\%) also from secondary vertices in dead material of the tracking
detector.

\begin{figure}[t]
\begin{center}
\includegraphics[width=0.4\textwidth,height=10cm,trim=0 50 0 15]
{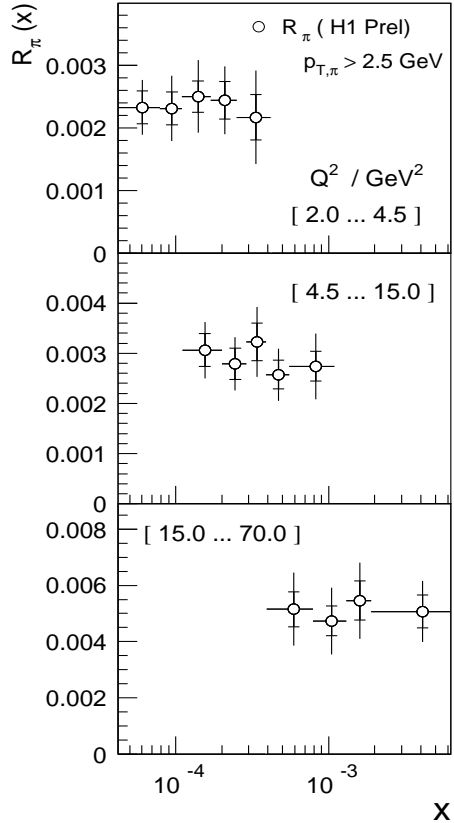}
\end{center}
\caption{\it The rate  of {\piz}-meson production in DIS as a 
function of {\xbj}
  obtained by dividing the cross section shown in Figure (a) by the
  inclusive ep cross-section in each bin of {\xbj},{\ybj} and {\qsq}.}
\label{fig:raxbj}
\end{figure}

\section{RESULTS}

The final experimental
results of the analysis are obtained as differential cross sections of
forward {\piz}-meson production as a function of {\qsq}, and 
as a function of {\xbj},
{\etapi} and {\ptpi} in three regions of {\qsq} for {\ptpi} 
$> 2.5$~GeV.
 In addition the {\piz} cross sections as a function of {\xbj}
and {\qsq} are measured for {\ptpi}~$>$ 3.5~GeV.
The phase space is given by 0.1 $<$ {\ybj} $<$
0.6, 2 $<$ {\qsq} $<$ 70~{\gevsq}, $5^\circ <$ {\thpi} 
$< 25^\circ$ and {\xpi} $=$ {\epi}$/${\eprot} $>$ 0.01 in addition to
the {\ptpi} thresholds given above. {\thpi} and {\xpi} are taken in
the H1 laboratory frame, {\ptpi} is calculated in the hadronic cms.
The measurement extends down to values of 
{\xbj} $>$ 5$\cdot$10$^{-5}$, covering two orders of magnitude in {\xbj}.
 Of these cross sections~\cite{h1pi096} only
$d{\sigma_{\pi}}/d${\xbj} for {\ptpi} $> 2.5$~GeV are shown here.
All observables are corrected for detector effects and for the
influence of QED radiation by a bin-by-bin unfolding procedure.
The typical systematic uncertainty is 15-25$\%$, compared to a
statistical uncertainty of about 10$\%$. Contributions to the
systematic error include among others the energy scales of the
calorimeter, uncertainties in the selection of {\piz}-mesons and 
the model dependence of the
bin-by-bin correction procedure.

The cross sections as a function of {\xbj} shown in Figure
\ref{fig:xbj} exhibit a strong rise towards small {\xbj}. An
interesting observation is that this rise corresponds to the rise of
the inclusive ep cross section. The ratio of the two cross sections is
shown in Figure \ref{fig:raxbj} and shows no dependence on {\xbj}
in the three regions of {\qsq}. The production 
rates do decrease with decreasing $Q^2$ for fixed $x$. The inclusive
ep cross section for this comparison is obtained by integrating the H1
QCD fit to the 1996 structure function data as presented
in~\cite{h1f2jer} for every bin of the present measurement of
inclusive {\piz}-meson cross sections.

The DGLAP prediction for pointlike virtual photon scattering
only, represented by LEPTO~\cite{lepto}, falls
clearly below the data. 
The mechanism of emitting partons according to the DGLAP
splitting functions, combined with pointlike virtual 
photon scattering only, is clearly not supported by
the data in particular at low $x$. 
A considerable improvement of the description of the data
is achieved by considering processes where
the virtual photon entering the scattering process is resolved. Such a
prediction is provided by RAPGAP~\cite{rapgap}. 
 All predicted cross sections increase by up
to 30$\%$ when the scale in the hard scattering is increased to $Q^{2} +
4p_{T}^{2}$ from $Q^{2} + p_{T}^{2}$~\cite{twthesis}, and hence does
not improve the overall 
description.
Whether this mechanism is adequate to describe the 
{\piz} cross sections down to the lowest available $x$ therefore cannot
 finally be decided by RAPGAP.

Next we compare with a 
 calculation following the BFKL formalism 
in order ${\cal O}(\alpha_s)$. Fragmentation
functions are used to calculate the {\piz}-meson cross sections from
the partonic final state.
The predictions obtained with these calculations turn out to be in good
agreement with the neutral pion cross sections  measured in the entire
available phase space with a slight tendency to be below the data at
the lowest values of $x$ available.

\section{CONCLUSIONS}
With the present measurement of inclusive {\piz}-meson cross sections
it has become possible for the first time 
to measure observables of the hadronic final state in this
region of the phase space with relatively small experimental
uncertainties. It provides testing ground for any theory that claims
to describe processes at small {\xbj} with large phase space for 
parton emissions and a reasonably hard scale.

Models using ${\cal O}(\alpha_s)$
QCD matrix elements and parton cascades according to the DGLAP
splitting functions cannot describe the differential neutral pion
cross sections at low $x$. Including processes in which the virtual
photon is resolved leads to an improved description of the data.
Renormalization and factorization scale uncertainties however limit
the precision of the predictions. A
calculation based on the  BFKL formalism is in good agreement with the
data. Considering the relatively small uncertainties~\cite{KwMaOu}
given for this
calculation it is the best available approximation of QCD in
the considered phase space.

\begin{small}

\end{small}

\end{document}